\documentclass[aps,prd,showpacs,preprint,amsmath,amssymb,nofootinbib]{revtex4-1}
\usepackage{graphics} %figure files
\usepackage{graphicx,color}
\usepackage{epsfig} %eps figure
\usepackage{longtable}
\usepackage{bm}
\usepackage{color}

\begin{document}

\title{Nonextensive Kinetic Theory and H-Theorem in General Relativity}

\author{A. P. Santos$^{1}$}
\email{alysonpaulos@gmail.com}

\author{R. Silva$^{2,3}$}
\email{raimundosilva@fisica.ufrn.br}

\author{J. S. Alcaniz$^{4}$}
\email{alcaniz@on.br}

\author{J. A. S. Lima$^{5}$}
\email{jas.lima@iag.usp.br}

\affiliation{$^{1}$ Instituto de Ciencias Exatas e Tecnologia, Universidade Federal do Amazonas, Itacoatiara-AM, Brasil}
\affiliation{$^{2}$Universidade Federal do Rio Grande do Norte,
Departamento de F\'{\i}sica, Natal - RN, 59072-970, Brazil}
\affiliation{$^{3}$ Departamento de F\'{\i}sica, Universidade do Estado do Rio Grande do Norte, Mossor\'o, 59610-210, Brasil}
\affiliation{$^{4}$ Departamento de Astronomia, Observat\'orio Nacional, 20921-400 Rio de Janeiro RJ, Brasil}
\affiliation{$^{5}$ Departamento de Astronomia (IAG-USP), Universidade de S\~ao Paulo, 05508-090 São Paulo SP,  Brasil}


\date{\today}

\begin{abstract}
The nonextensive kinetic theory for degenerate quantum gases is discussed in the general relativistic framework.  By incorporating nonadditive modifications in the collisional term of the relativistic Boltzmann equation and entropy current, it is shown that Tsallis entropic framework satisfies a $H$-theorem  in the presence of gravitational fields.  Consistency with the 2nd law of thermodynamics is obtained only whether the entropic $q$-parameter lies in the interval  $q\in[0,2]$. As occurs in the absence of gravitational fields, it is also proved that the local collisional equilibrium is described by the extended Bose-Einstein (Fermi-Dirac) $q$-distributions.	
\end{abstract}



\maketitle

Some extensions of the orthodox Boltzmann-Gibbs-Shannon (BGS) entropy have been proposed in order to address the statistical and kinetic behavior {of anomalous nonadditive} systems \cite{tsallisbook,kaniadakisreview}, among them, the Tsallis \cite{tsallis88} and Kaniadakis \cite{kaniadakis2002} statistics (see also \cite{entropies} for other possibilities).

Here we focus our attention on the nonextensive Tsallis entropy which for a classical non-degenerated gas system of point particles reads      (unless explicitly stated, in our units $k_B = c=1$)
\begin{equation}\label{Sq}
S_q\, = -\int f^{q}\ln_q f d^3p, \,\,\,\,\,
\end{equation}
where $q$ is a real number quantifying the degree of nonadditivity of $S_q$, and $\ln_q(f)$ is the nonadditive $q$-logarithmic function whose  inverse is the $q$-exponential. Both functions are defined by:
\begin{eqnarray}\label{E1}
\ln_q(f)=\,(1-q)^{-1}(f^{1-q}-1),\,\,\, (f>0),\\
e_q(f) = [1 + (1-q)f]^{\frac{1}{1-q}},\,\,\,e_q(\ln_qf)=f,
\end{eqnarray}
which reduce to the standard expressions in the limit $q \rightarrow 1$. The above formulas also imply that for a gas system composed by two subsystems (A,B), the kinetic Tsallis measure verifies $S_q(A + B)= S_q(A) + S_q(B) + (1-q)S_q(A)S_q(B)$. Hence, for $q=1$  the logarithm extensive measure associated to the GBS approach is recovered.

The validity  of the 2nd law of thermodynamics for the nonextensive path was kinetically investigated in the classical \cite{prl2001}, special relativistic \cite{pre2005} and quantum-mechanical domains \cite{epl2010,abePRL2003} (see also \cite{silva98} for a simplified kinetic derivation of the q-Maxwellian based on a nonadditive extension of the ansatz adopted by Maxwell in his seminal paper \cite{Max1860}). {The thermodynamic  consistency of all} these short-range regimes provided the constraint $q\in[0,2]$, which was also obtained based on the convexity property of the extended relativistic entropy in the quantum regime \cite{abePRL2003} (cf. \cite{bentoPRE2015} for tighter bounds based on the third law of thermodynamics).

More recently, relaxing the local Maxwellian equilibrium distribution, the classical  Chandrashekar result for dynamical friction in Newtonian gravity was investigated in the $q$-nonextensive context. It was found that the large timescale of globular clusters  spiraling to the center of the Milky Way, as obtained from N-body simulations, can be interpreted as a departure from the extensive Gaussian result in agreement  with Tsallis power-law approach \cite{Silva2016}.

On the other hand, we recall that the foundations of the kinetic counterpart of BGS statistics in the framework of general relativity theory (GRT) has been investigated and understood since long ago \cite{stwart,israel,ehlers,weinberg}.
{In the following decades,}  this approach was applied to  relevant problems in astrophysics and cosmology, like stellar and cosmological nucleossynthesis, baryogenesis, and fluctuations of  the cosmic background radiation \cite{KT90,JB89,Dod03,Weinberg08}).  However, as far as we know, the same does not occurs with the nonextensive Tsallis proposal. In principle, this is a clear obstacle to enlarge the field of applications for testing the nonextensive kinetic formalism which is considered more convenient to systems endowed with long range forces.

In this letter we fill this gap providing an analytical proof of the $H$-theorem for degenerate nonextensive quantum gases in curved spacetimes.  To this end, we combine two new ingredients: (i) nonadditive entropy current, and (ii) possible effects of statistical correlations on the collisional term of the general relativistic kinetic equation ($q$-extension of the Boltzmann ``Stozssahlansatz'' hypothesis).  As we shall see,  the second law of thermodynamics (the positiveness of the source of entropy) also constrains the nonextensive parameter on the interval $0 \leq q \leq 2$. As an extra bonus, the results for the nonextensive general relativistic Juttner gas are also obtained as a particular case.
%\noindent {\it Liouville Operator and Transport Equation}.

In order to understand how nonextensive effects  alters the standard approach, let us now consider a relativistic rarified quantum gas in a curved space-time which is out but not too far from equilibrium.  The gas particles are supposed to interact through short and long-range forces.  The former act only during the instantaneous binary collisions while the latter are described by the metric tensor $g_{\mu\nu}$. Following standard lines, the evolution of the one particle distribution function in phase space, $f(x,p)\equiv f(x^{\mu}, p^{\mu})$, is governed by the kinetic Boltzmann equation in the presence of gravitation \cite{stwart,israel,weinberg,ehlers}
\begin{equation}\label{stw7}
L[f]\equiv p^\mu\frac{\partial f}{\partial x^\mu}-\Gamma^\mu_{\nu\alpha}p^\nu
p^\alpha\frac{\partial f}{\partial p^\mu}=C[f]\,,
\end{equation}
where $L[f]$ is the Liouville operator and $C[f]$ is the invariant phase space density of collisions. In the above equation is also implicit that $f(x,p)p^\mu u_\mu d\Sigma dP$ is the particle number whose world lines intersect the volume element $u_\mu d\Sigma$ around $x$ and a 4-momentum defined in the range $(p, p+dp)$, such that $\mu, \nu, \beta, ...= 0, 1, 2, 3$. As the quantum gas has particles with rest mass $m$, their 4-momentum must be restricted the part of the space of moments, therefore an element of volume for this portion of phase space is defined by
\begin{equation}\label{stw4}
dP:=A(p)\delta(p_\mu p^\mu+m^2)\sqrt{-g}dp^0dp^1dp^2dp^3\,,
\end{equation}
{ where $A(p)=2$ if $p^0>0$, and $A(p)=0$ for all other cases.} The adopted signature is (-,+,+,+) so that  $p_\mu p^\mu = -m^{2}$.

{Following standard lines, the unified invariant collisional term (for all three different statistics)} can be written as
\begin{widetext}
\begin{equation}\label{stw8}
C[f]=\int\int\int[\hat{f}\hat{f}_1f^{'}f^{'}_1-ff_1\hat{f}^{'}\hat{f}^{'}_1]
W(pp_1\rightarrow p^{'}p^{'}_1)dP_1dP^{'}dP^{'}_1\,.
\end{equation}
\end{widetext}
Here, the particle distribution is given by $\hat{f} \equiv 1 +\kappa\frac{h^3}{g_s} f$, where $h$ is the Planck constant and $g_s$ is the degeneracy factor of particles with spin $s$. In particular, the number  $\kappa = +1$ refers to bosons (due to the statistics of indistinguishable particles),  $\kappa = -1$ are fermions (due to Pauli exclusion principle), while $\kappa = 0$ stand for particles without spin. The scalar collision frequency, $W(pp_1\rightarrow p^{'}p^{'}_1)$, depends only of the momentum variables and remains invariant under space and time reflexions, $W(pp_1\rightarrow p^{'}p^{'}_1)=W(p^{'}p^{'}_1\rightarrow pp_1)$.
%\cite{Dolgov}.

The macroscopic variables and the balance equations describing the thermodynamic states of the gas particles are microscopically calculated through the relativistic distribution function $f(x,p)$, which here also depends on the $\kappa$ parameter. The key macroscopic variable to the proof of the H-theorem is the entropy current defined by
\begin{equation}\label{stw9}
S^\mu(x)=-\int p^\mu[f\ln f -\frac{g_s}{\kappa h^3}  \hat{f}\ln\hat{f}]dP,
\end{equation}
from which one obtains an expression for the second law of thermodynamics
\begin{equation}\label{stw10}
\nabla_\mu S^\mu = -\int \ln (\frac{f}{\hat f})L(f)dP \geq 0,
\end{equation}
where $\nabla_\mu$ denotes the covariant derivative.

By rewritten  the above expression in terms of \eqref{stw8}, one may check that the r.h.s. of the above equality is necessarily nonnegative thereby proving the standard H-theorem \cite{stwart}.  The collisional equilibrium states are obtained  when $\ln (f/\hat {f})$ is a summational invariant. In this case, the integral vanishes identically and the unified form of the equilibrium  distribution function is derived
\begin{equation}\label{cov1}
f_0(x,p) = \frac{g_s}{h^3}[\exp(\tilde{\alpha}-\beta_\mu p^\mu) - \kappa]^{-1},
\end{equation}
where $\tilde{\alpha}=\ln(g_s/h^3)-\alpha$. Note that for $\kappa=0$ the  relativistic form of the Maxwell-Boltzmann-Juttner distribution is recovered.

At this point, it is worth mentioning that the collisional operator \eqref{stw8} is based on the Boltzmann``Stossahlansatz''  hypothesis, which means that colliding particles are uncorrelated. This celebrated nonmechanical  assumption implies that the two point correlation functions are factorizable [$f(x,p,p_1) = f(x,p)f(x,p_1)$].

Let us now discuss the impact of the nonextensive entropic approach in the covariant kinetic theory, and how the  $H_q$-theorem for a quantum dilute gas is harmonized with the curved spacetime description.  In principle,  the Boltzmann ``Stossahlansatz", as well as the own measure of entropy will be modified \cite{prl2001,pre2005,epl2010}. The new hypothesis  combined with changes in the entropy flux and collisional term, encode  all the  statistical properties of the nonextensive microscopic description of a dilute quantum gas.

To begin with,  we propose to the entropy current (\ref{stw9}) the following nonadditive expression:
\begin{equation}\label{cov2}
S^\mu_q(x)=-\int p^\mu[f^q\ln_qf - \frac{g_s}{\kappa h^3}\hat{f}^q\ln_q\hat{f}]dP,
\end{equation}
where the function $\ln_q$ has been defined in \eqref{E1}, and $\hat{f}^q=( 1 +\kappa\frac{h^3}{g_s} f)^q$ [see definition below \eqref{stw8}]. Operationally, it means that not only the logarithm functions are replaced by power laws (q-ln), but also that the collisional term must be suitably modified.

By introducing the useful notation
\begin{equation}\label{cov5}
G_q(f) \equiv f^q\ln_qf -\frac{g_s}{\kappa h^3} \hat{f}^q\ln_q\hat{f},
\end{equation}
it is easy to check that
\begin{equation}\label{cov4}
\nabla_\mu S_q^\mu=-\int G_q^{'}(f)L[f]dP,
\end{equation}
where $G_q^{'}(f)$ is the derivative with respect to $f$ which is given by
\begin{equation}\label{cov6}
G_q^{'}(f)=q[f^{q-1}\ln_qf - \hat{f}^{q-1}\ln_q\hat{f}].
\end{equation}
and using the duality property $ q^*\rightarrow 2-q$, the above expression becomes
\begin{equation}\label{cov7}
G_q^{'}(f)=q[\ln_{q^{*}}f - \ln_{q^{*}}\hat{f}].
%\nonumber
\end{equation}
At this point it is interesting to consider two basic identities defining a deformed algebra involving q-logarithm and q-exponential functions  which will be used to simplify several expressions. The q-difference and q-product is defined by \cite{borges}:

\begin{subequations}
\begin{equation} \label{qdif}
x \ominus_{q*} y := \frac{x-y}{1+(1-q^*)y} \ \ \ \ \forall \ \ y \neq \frac{1}{1-q^*},
\nonumber
\end{equation}
\begin{equation} \label{qprod}
x \otimes_{q*} y := \left[x^{1-q_*} + y^{1-q_*} - 1 \right]^{\frac{1}{1-q_*}} \ \ \ \ x, y >0,
\nonumber
\end{equation}
\end{subequations}
which allow to write the q-ln and q-exp in a more compact form:
\begin{subequations}\label{cov8}
\begin{equation}
\ln_{q^{*}}(f/\hat{f})\equiv\ln_{q^{*}}(f)\ominus_{q^{*}}\ln_{q^{*}}(\hat{f})=
\frac{\ln_{q^{*}}(f)-\ln_{q^{*}}(\hat{f})}{1+(1-q^*)\ln_{q^*}\hat{f}},
\end{equation}
%\be
%\ln_{q^{*}}(f/\hat{f})\equiv\ln_{q^{*}}(f)\ominus_{q^{*}}\ln_{q^{*}}(\hat{f})=
%\frac{\ln_{q^{*}}(f)-\ln_{q^{*}}(\hat{f})}{1-(1-q^*)\ln_{q^*}\hat{f}},
%\ee
\end{subequations}

implies that (\ref{cov6}) can be recast as
\begin{equation}\label{cov9}
G_q^{'}(f)=q\hat{f}^{1-q^*}\ln_{q^{*}}(f/\hat{f}),
\end{equation}
with expression~(\ref{cov4}) taking the form
\begin{equation}\label{cov9a}
\nabla_\mu S_q^\mu=-q\int\hat{f}^{1-q^*}\ln_{q^{*}}(f/\hat{f})L[f]dP,
\end{equation}
which should be compared with (\ref{stw10}). Actually,  all the above  expressions reduce to the standard ones in the mathematical limit $q, q^{*} \rightarrow 1$.

On the other hand, nonextensive effects can be incorporated in the general relativistic Boltzmann equation (\ref{stw7})  only through the collisonal term  because the Liouville operator is not modified. Therefore, let us consider the following transport equation
\begin{equation}
\label{BEQ}
L[f]=C_q[f],
\end{equation}
where beyond the standard requirements the structure of $C_q(f)$ does not obey the ``Stossahlansatz" and $\lim_{q \rightarrow 1} C_q(f) = C(f)$ \cite{prl2001,pre2005}. It is natural to consider that Boltzmann's ``Stosshalansatz" is not valid because the nonextensive effects give rise to statistical correlations requiring an extension  of the molecular chaos hypothesis, an assumption underlying the Boltzmann approach which still remains controversial nowadays \cite{Zee92}. In order to capture correctly the effects of gravitational field in what follows we consider the following expression:
\begin{eqnarray}\label{cov10}
C_q[f]&=&\int\int\int\hat{f}^{q^*}\hat{f}_1\hat{f}^{'}\hat{f}^{'}_1
\left[\left(\frac{f^{'}}{\hat{f}^{'}}\right)\otimes_{q^*}
\left(\frac{f_1^{'}}{\hat{f}_1^{'}}\right)\right. \nonumber\\
& & \left. -\left(\frac{f}{\hat{f}}\right)
\otimes_{q^*}\left(\frac{f_1}{\hat{f}_1}\right)\right]WdP_1dP^{'}dP^{'}_1.
\end{eqnarray}
which reduces to  (\ref{stw8}) in the appropriate limit.

Once the basic nonextensive modifications on the 4-entropy and collisional term have been explicited the proof of the covariant version of the $H_q$-theorem ($\nabla_\mu S_q^\mu \geq 0$) in the context of general relativity can be performed.  By replacing the equation~(\ref{cov10}) in (\ref{cov9a}), we obtain
\begin{eqnarray}\label{cov11}
\nabla_\mu S_q^\mu &=& q\int\int\int\hat{f}\hat{f}_1\hat{f}^{'}\hat{f}^{'}_1\
ln_{q^*}\left(\frac{f}{\hat{f}}\right)
\left[\left(\frac{f^{'}}{\hat{f}^{'}}\right)\otimes_{q^*}
\left(\frac{f_1^{'}}{\hat{f}_1^{'}}\right)\right. \nonumber\\
& & \left. -\left(\frac{f}{\hat{f}}\right)
\otimes_{q^*}\left(\frac{f_1}{\hat{f}_1}\right)\right]WdP_1dP^{'}dP^{'}_1.
\end{eqnarray}
By using the invariance of term $W(pp_1\rightarrow p^{'}p^{'}_1)$ in
the above relation, as well as by considering the permutations of the type $(p\rightarrow
p_1)\Rightarrow(f\rightarrow f_1)$,
$(p\rightarrow p^{'})\Rightarrow(f\rightarrow f^{'})$ and
$(p_1\rightarrow p_1^{'})\Rightarrow(f_1\rightarrow f_1^{'})$, it is possible to show that

\begin{eqnarray}
&&\nabla_\mu S_q^\mu = \frac{q}{4}\int\int\int\hat{f}\hat{f}_1
\hat{f}^{'}\hat{f}^{'}_1\times \\ %\nonumber\\
& & \left[ \ln_{q^*}\left(\frac{f}{\hat{f}}\right)+
\ln_{q^*}\left(\frac{f_1}{\hat{f}_1}\right)-\ln_{q^*}
\left(\frac{f^{'}}{\hat{f}^{'}}\right)-\ln_{q^*}\left(
\frac{f_1^{'}}{\hat{f}_1^{'}}\right)\right]
\nonumber\\
& & \left[\left(\frac{f^{'}}{\hat{f}^{'}}\right)\otimes_{q^*}
\left(\frac{f_1^{'}}{\hat{f}_1^{'}}\right)-\left(
\frac{f}{\hat{f}}\right)\otimes_{q^*}\left(\frac{f_1}{\hat{f}_1}
\right)\right]WdP_1dP^{'}dP^{'}_1.\nonumber\\
& & \nonumber
\end{eqnarray}
Now, in order to simplify the above expression, let us use the following identity based upon extended algebra, i.e $\ln_{q^*}(a\otimes_{q^*}b)=
\ln_{q^*}(a)+\ln_{q^*}(b)$. Thereby, we obtain
\begin{widetext}
\begin{equation}\label{cov12}
\nabla_\mu S_q^\mu = \tau_q =
\frac{q}{4}\int\int\int\hat{f}\hat{f}_1
\hat{f}^{'}\hat{f}^{'}_1[\ln_{q^*}Z-\ln_{q^*}Y](Z-Y)WdP_1dP^{'}dP^{'}_1,
\end{equation}
\end{widetext}
where
\begin{equation}
Z=\left(\frac{f^{'}}{\hat{f}^{'}}\right)\otimes_{q^*}
\left(\frac{f_1^{'}}{\hat{f}_1^{'}}\right),\quad\quad Y=
\left(\frac{f}{\hat{f}}\right)\otimes_{q^*}
\left(\frac{f_1}{\hat{f}_1}\right),
\end{equation}
and $\tau_q$ is the entropy source which must be positive due the second law of thermodynamics. This positiveness of the entropy source is related with the inequality $(Z-Y)[\ln_{q^*}Z-\ln_{q^*}Y]\geq0$. It shows that the generally covariant $H$-theorem is alo consistent with nonadditive effects present in the Tsallis framework because
\begin{equation}\label{cov12}
\nabla_\mu S_q^\mu=\tau_q\geq0.
\end{equation}
Note that, for, $q<0$ or $q>2$, the equation~(\ref{cov12}) is a decreasing function of time, being thermodynamically forbidden. It thus follows that the parameter $q$ is kinetically restricted on the interval $[0,2]$. Such constraint is compatible with those which have been calculated through the $H_q$-theorem in the classical, relativistic and quantum-mechanics regimes
\cite{prl2001,pre2005,epl2010}, as well as through an approach of quantum Clausius's inequality~\cite{abe03}. The inequality (\ref{cov12}) shows that the q-entropy source must be
positive or zero thereby furnishing a kinetic argument for
the second law of thermodynamics in the nonextensive formalism. Naturally,  like in the standard Boltzmann approach,  this argument
does not provide a complete kinetic proof of the second law since an extra statistical assumption is also required.
By imposing that the source of entropy vanishes ($\tau_q=0$) we obtain the
necessary and sufficient condition stablishing the local equilibrium states. It also implies that  $L(f)\equiv 0$ which means  that $f$ is a constant of motion.  The simplest case happens in the absence of collisions ($W=0$). In the collisional equilibrium case, the distribution function follows from the condition of detailed balance:
\begin{equation}\label{stw18}
\ln_{q^*}\left(\frac{f}{\hat{f}}\right)+\ln_{q^*}\left(\frac{f_1}{\hat{f}_1}\right)=
\ln_{q^*}\left(\frac{f^{'}}{\hat{f}^{'}}\right)+\ln_{q^*}\left(\frac{f^{'}_1}{\hat{f}^{'}_1}\right).
\end{equation}
The above sum of $ln_q$-terms  remains constant during a collision (invariant summational).  It is well known that the most general invariant collisional  is a linear combination involving a scalar and the 4-momentum $p^\mu$ (see for instance \cite{ehlers}). Consequently, the distribution function is given by the condition
\begin{equation}\label{cov13}
\ln_{q^*}\left(\frac{f}{\hat{f}}\right)=\alpha+\beta_\mu p^\mu,
\end{equation}
where $\alpha(x)$ and $\beta_\mu=\beta_\mu(x)$ are quantities conserved in the collisions.
By solving for $f$ we obtain

\begin{equation}\label{cov14}
f=\frac{g_s}{h^3}\left[\frac{h^3}{g_s}\exp_{q^*}\left(-\alpha(x)-\beta_\mu p^\mu(x)\right) - \kappa\right]^{-1},
\end{equation}

It is possible to show that the moments from $f$ are finite whether $\beta_\mu(x)$ is a time-like vector directed for the future, i.e., $\beta_\mu(x)=\beta(x)u_\mu$, where $u_\mu u^\mu=-1$, $u^0>0$ and $\beta>0$.
By replacing (\ref{cov14}) into Boltzmann's equation (\ref{BEQ}) with vanishing collision integral, it follows that
\begin{equation}
\partial_\mu \alpha(x)p^\mu+\nabla_\nu(\beta_\gamma)p^\gamma p^\nu=0.
\end{equation}
If $m\neq0$ and $\beta_\mu$ is along the geodesic { and $p^\mu$ is parallel
transported}, we have
\begin{equation}
\nabla_\nu(\beta_\gamma)=0,
\end{equation}
and therefore $\alpha$ is a constant. $\beta_\mu$ is a constant as well as a time-like vector, thus we obtain the expressions

\begin{equation}\label{cov16}
f=\frac{g_s}{h^3}\left[\exp_{q^*}(\tilde{\alpha}-\tilde{\beta_\mu} p^\mu) - \kappa\right]^{-1},
\end{equation}
where $\tilde{\alpha}=\ln_{q^*}(g_s/h^3) - (g_s/h^3)^{q^*-1}\alpha$ and $\tilde{\beta_\mu}=(g_s/h^3)^{q^*-1}\beta_\mu$, which reduces to the extensive result in the limit $q, q* \rightarrow 1$ (cf. \eqref{cov1}).

For $\kappa=\pm 1$, the above expression  represents the Bose-Einstein and Fermi-Dirac relativistic distributions in the Tsallis statistical formalism. When quantum effects are negligible  ($\kappa = 0$), it reduces to the relativistic nonextensive distribution \cite{pre2005}
\begin{equation}
f_0(x,p)=\frac{g}{h^3}\exp_{q^*}(\tilde{\alpha}-\tilde{\beta}_\mu p^\mu),
\end{equation}
Finally, in the nonrelativistic limit the nonextensive Maxwell distribution, which is the basis of nonextensive statistical mechanics is properly calculated \cite{silva98}. In particular, in the extensive (or additive entropy) limit $q\rightarrow 1$ the standard cases are recovered.

Broadly speaking, the relativistic nonextensive $q$-distribution in the presence of gravitation is the most general approach when nonadditive effects upon the statistical physics are taken into account. Special cases like the special relativistic, classical and quantum limits \cite{prl2001,pre2005,epl2010} are all covered here. Even the nonrelativistic distribution is also obtained as a limiting case of expression (\ref{cov16}).  Physically, it is also very compelling that the heart of the H-theorem, that is, the fact that the entropy source is non-negative, as required by the second law of thermodynamics, constrains the  nonextensive q-parameter on the same interval $[0,2]$ irrespective of the physical regime \cite{prl2001,pre2005,epl2010}.

It should be also stressed  that a simplifying  and more direct deduction could be obtained by starting with the Boltzmann equation in the comoving frame. This argument provides a more direct proof of $H_q$-theorem. However, the approach based on the $q$-calculus and $q$-algebra as discussed in Ref. [22] and adopted here, may become an useful tool for different problems involving  the q-statistical framework.

Furthermore, there are trivial and nontrivial kinetic effects in the general relativistic (GR) framework. For instance, in the calculations of the transport coefficients \cite{israel,W71} the local expressions are equivalent in both special relativity and GR context. Also, the same must be true for the $q$-relativistic kinetic approach. However, non-trivial GR kinetic effects are also ordinarily expected in several situations, as for instance, in the cosmological domain. In this concern, current calculations of the primordial power spectrum of matter perturbations and the temperature anisotropies of the cosmic background radiation are calculated through the Boltzmann equation and nontrivial effects appear there due to  specific gravity contributions coming from the local relativistic gravitational potential \cite{Dod03}. Indeed, the same must happen when the relativistic q-distribution is adopted. In principle, it should be interesting to investigate how the q-kinetic approach as discussed here modify the standard results.

Last but not least,  the nonextensive general relativistic distribution (\ref{cov14}) derived here must also be valid in the context of more general metric gravitational theories.

\begin{acknowledgements}
APS  wishes to thank FAPEAM (Funda\c c\~ao de Apoio a Pesquisa do Amazonas) for the DCR fellowship. This work was also partially supported by the Brazilian agency CNPq, CAPES (PROCAD 2013), FAPERJ (Funda\c c\~ao de Apoio a Pesquisa do Estado do Rio de Janeiro) and FAPESP (Funda\c c\~ao de Apoio a Pesquisa do Estado de S\~ao Paulo).		
\end{acknowledgements}

\end{document}